\documentclass{ws-procs9x6-cpt16}
\usepackage{xspace}

\begin{document}

\newcommand{\refeq}[1]{(\ref{#1})}
\def\etal {{\it et al.}}

\def\ksn{\ensuremath{{K_S}}\xspace}
\def\kln{\ensuremath{{K_L}}\xspace}
\def\kn{\ensuremath{{K^0}}\xspace}
\def\knb{\ensuremath{{\bar{K}^0}}\xspace}

\title{Search for CPT and Lorentz-Symmetry Violation in \\
Entangled Neutral Kaons}

\author{Antonio Di Domenico}

\address{Dipartimento di Fisica, Sapienza Universit\`a di Roma and
INFN Sezione di Roma \\
P.\ le A.\ Moro 2, I-00185, Rome, Italy}

\begin{abstract}
The neutral-kaon system constitutes a fantastic and unique laboratory
for the study of CPT symmetry and the basic principles of quantum mechanics, 
and a $\phi$-factory represents a unique opportunity to push forward
these studies.
The experimental results show no deviation from the expectations 
of quantum mechanics and CPT symmetry, while the extreme precision 
of the measurements, in some cases, 
reaches the interesting Planck-scale region. 
At present the KLOE-2 experiment is collecting data 
with an upgraded detector with the aim of significantly 
improving these kinds of experimental tests. 
\end{abstract}

\bodymatter

\section{Introduction}

A violation of CPT symmetry would have a dramatic impact on our present theoretical picture and would definitely constitute an unambiguous signal of a new physics framework, thus strongly motivating both experimental searches and theoretical studies on this subject.
In attempts to discuss quantum-gravity scenarios, 
speculative theoretical models have been considered which 
may exhibit a CPT-symmetry breakdown.\cite{mavroreview,liberatireview} 
Among them
a general theoretical possibility for CPT violation 
is provided by the Standard-Model Extension (SME),
based on spontaneous breaking of Lorentz symmetry,\cite{kost1}
which appears to be compatible with the basic
tenets of quantum field theory and retains gauge
invariance and renormalizability.
 
The neutral-kaon doublet is one of the most intriguing systems in nature.
During its time evolution a neutral kaon oscillates between its particle and 
antiparticle states with a beat frequency 
$\Delta m 
\approx 5 \times 10^9 
\hbox{ s}^{-1}$
($\approx 3 \times 10^{-15} \hbox{~GeV}$),
where $\Delta m$ 
is the tiny mass difference between the two
physical states $K_L$ and $K_S$,
exponentially decaying with very different lifetimes, $\tau_L \gg \tau_S$.
The fortunate coincidence that 
$\Delta m$
is about half the 
decay width of $K_S$ 
allows observing a variety 
of intricate quantum interference phenomena 
in the time evolution and decay of neutral kaons. 

At a $\phi$-factory
neutral-kaon pairs are produced 
in a pure antisymmetric entangled state, offering new and unique possibilities 
to study 
the discrete symmetries and the basic principles of quantum mechanics.\cite{01}
What makes the entangled $\kn\knb$ pair a really unique system, even with respect to 
other similar neutral-meson systems ($B^0_d$, $B^0_s$, and $D^0$),
is the presence of peculiar and strong amplification mechanisms in the 
CPT-violation observables.
At a $\phi$-factory
the precision of the measurements in some cases 
can  reach the 
level of the interesting Planck-scale region,
i.e., $\mathcal {O} (m^2_K/M_{Planck}) \sim 2 \times 10^{-20} \,\mbox{GeV}$,
which is a very remarkable level of accuracy.

\section{``Standard" CPT test from unitarity}
 
The complex parameter $\delta$ describes CPT violation
 in $\kn$-$\knb$ mixing, and it is proportional to the particle-antiparticle mass and width difference:
\begin{equation}
\delta = \tfrac{1}{2}\frac{(m_{\knb}-m_{\kn})-i(\Gamma_{\knb}-\Gamma_{\kn})/2}{\Delta m +i\Delta\Gamma/2}~.
\label{eq:delta}
\end{equation}
 The real part of $\delta$ was 
 measured by CPLEAR\cite{cplearred} studying the time behaviour of semileptonic decays
from initially tagged \kn and \knb mesons, while the imaginary part can be bounded imposing the unitarity 
condition.\cite{BS,kloebs,ktevepsp}

    The limits on $\delta$
can be used
   to constrain the mass and width difference between \kn
   and \knb.
For $\Gamma_{\kn}-\Gamma_{\knb}=0$, i.e., 
neglecting CPT-violating effects in the decay amplitudes,
Eq.\ (\ref{eq:delta}) translates into the best bound on the 
fractional mass difference:
$\left|{m_{\kn}-m_{\knb}}\right|< 4.0  \times 10^{-19}$ GeV 
at 95\%~C.L.
It is worth noting that this stringent limit is obtained thanks to the amplifying effect of the denominator in Eq.\ (\ref{eq:delta}), due to
 the tiny mass and width difference between the physical states \ksn and \kln.

\section{CPT- and Lorentz-symmetry tests}

In the SME for neutral kaons, CPT violation manifests to lowest order 
only in the mixing parameter 
$\delta$, (e.g., vanishes at first order in the decay amplitudes), 
and exhibits a dependence on the 4-momentum of the kaon:
\begin{eqnarray}
\label{eq:deltak}
\delta \approx i \sin \phi_{SW} e^{i \phi_{SW}} \gamma_K (\Delta a_0-
\vec{\beta_K}\cdot \Delta{\vec{a}})/\Delta m,
\end{eqnarray}  
where $\gamma_K$ and $\vec{\beta_K}$ are the kaon boost factor and velocity in the observer frame, 
$\phi_{SW}$ is the so called {\it superweak} phase, 
and $\Delta a_{\mu}$
are four CPT- and Lorentz-violating coefficients for the two valence quarks in the kaon.

By studying the interference pattern of the entangled neutral kaon pairs in the
$\phi\rightarrow\kn\knb\rightarrow\pi^+\pi^- \pi^+\pi^- $ final state, as a function of sidereal time and particle direction in
celestial coordinates, the KLOE collaboration obtained the following results:\cite{Babusci:2013gda}
\begin{eqnarray}
\Delta a_0 &=& (-6.0 \pm 7.7_{\rm stat} \pm 3.1_{\rm syst})\times 10^{-18}  {\rm ~GeV}, \nonumber\\
\Delta a_X &=& (\,\,\,0.9 \pm 1.5_{\rm stat} \pm 0.6_{\rm syst})\times 10^{-18} {\rm ~GeV}, \nonumber\\
\Delta a_Y &=& (-2.0 \pm 1.5_{\rm stat} \pm 0.5_{\rm syst})\times 10^{-18}  {\rm ~GeV}, \nonumber\\
\Delta a_Z &=& (\,\,\,3.1 \pm 1.7_{\rm stat} \pm 0.5_{\rm syst})\times 10^{-18} {\rm ~GeV} .
\end{eqnarray}
These results constitute the most sensitive measurements in the quark sector of the SME,
and can be compared to similar results
obtained in the $B$ and $D$ meson systems, where an accuracy 
of 
$\mathcal{O}(10^{-15}{\rm ~GeV})$ and $\mathcal{O}(10^{-13}{\rm ~GeV})$, respectively, has been reached.\cite{kost4,pippo1}

\section{Search for decoherence and CPT-violation effects}

The quantum interference between the two kaons initially in the 
entangled state 
and decaying in
the CP-violating channel $\phi\rightarrow \ksn\kln \rightarrow \pi^+\pi^-
\pi^+\pi^-$, has been observed for the first time by the KLOE collaboration.\cite{kloeqm,addfp}
The decoherence parameter has been measured: 
\begin{eqnarray}
\label{eq:deckloe2}
\zeta_{0\bar{0}}&=&(1.4\pm9.5_{\rm stat}\pm3.8_{\rm syst})\times 10^{-7}~,
\end{eqnarray}
compatible with the prediction of quantum mechanics $\zeta_{0\bar{0}}=0$ and no decoherence effect. 
This constitutes the most precise quantum coherence test
 for an entangled system, due to the peculiar
 CP-violation suppression present in this specific 
decay channel, which naturally amplifies the sensitivity of the decoherence effect.

A model for decoherence can be formulated\cite{ellis1}
in which 
neutral kaons
are
described by a 
density matrix $\rho$ that
obeys  a modified Liouville-von Neumann equation.
In this context $\gamma$ is one of the relevant parameters signalling decoherence and CPT violation.\cite{wald}
It 
has mass units
and in a quantum-gravity scenario it is presumed to be at most of
$\mathcal {O} (m^2_K/M_{Planck}) \sim 2 \times 10^{-20} \,\mbox{GeV}$.
The KLOE collaboration obtained the following result\cite{addfp}
compatible with no CPT violation:
\begin{eqnarray}
  \label{eq:kloegamma}
  \gamma &=& \left( 0.7 \pm 1.2_{\rm stat} \pm 0.3_{\rm syst} \right) \times 10^{-21}\, \mbox{GeV}~,
\end{eqnarray}
while the sensitivity reaches
the interesting region.

\section{Direct CPT test in transition processes}

 A novel CPT test has been recently studied in the neutral-kaon system based on
the direct comparison of a transition probability with
its 
CPT reverse transition.\cite{cpttrans} The appropriate preparation and detection of 
{\it in} and {\it out} states in both the reference and the reverse processes is 
made by exploiting the entanglement of neutral kaons produced in a
$\phi$-factory and using their decays as filtering measurements of the kaon
states.
The test can be easily implemented at KLOE and KLOE-2, while in the $B$-meson system a similar test has been performed.\cite{babartrans}

\section{Conclusions and perspectives}
The parameters related to several possible CPT 
violations effects, 
including decoherence and Lorentz-symmetry breaking effects which might be justified in a quantum-gravity framework, 
have been measured in the neutral-kaon system in some cases
with a precision that very interestingly reaches the
Planck scale region.

The KLOE physics program is continuing with the KLOE-2 experiment,
presently taking data at the DA$\Phi$NE facility with an upgraded detector.\cite{04}
Significant improvements are expected
in all these CPT tests.

\end{document}